\documentclass[10pt, a4paper, twocolumn, twoside]{article} 

\usepackage[english]{babel} 

\usepackage{microtype} 

\usepackage{amsmath,amsfonts,amsthm} 

\usepackage[svgnames]{xcolor} 

\usepackage[small, labelfont=bf, up]{caption} 

\usepackage{booktabs} 

\usepackage{lastpage} 

\usepackage{graphicx} 

\usepackage{enumitem} 
\setlist{noitemsep} 

\usepackage{sectsty} 
\allsectionsfont{\usefont{OT1}{phv}{b}{n}} 

\usepackage{geometry} 

\usepackage[T1]{fontenc} 
\usepackage[utf8]{inputenc} 

\usepackage{XCharter} 

\usepackage{hyperref}

\usepackage{journals}

\usepackage{fancyhdr} 
\newcommand{\shorttitle}[1]{\fancyhead[CE]{\textsl{#1}}}
\newcommand{\shortauthors}[1]{\fancyhead[CO]{\textsl{#1}}}

\fancypagestyle{firstpage}{ 
	\fancyhf{}
        \lhead{\vspace*{0.0em} \textit{\footnotesize 21$^{\rm st}$ European Workshop on White
            Dwarfs \\ Castanheira, Campos, Montgomery, eds. \\[-0.2em]
          July 23--27, 2018, Austin, Texas, USA}}
}

\date{}

\newcommand{\authorstyle}[1]{{\large\usefont{OT1}{phv}{b}{n}\color{DarkRed}#1}} 

\newcommand{\institution}[1]{{\footnotesize\usefont{OT1}{phv}{m}{sl}\color{Black}#1}} 

\usepackage{titling} 

\newcommand{\HorRule}{\color{DarkGoldenrod}\rule{\linewidth}{1pt}} 

\pretitle{
	\vspace{-30pt} 
	\HorRule\vspace{10pt} 
	\fontsize{16}{12}\usefont{OT1}{phv}{b}{n}\selectfont 
	\color{DarkRed} 
}

\posttitle{\par\vskip 10pt} 

\preauthor{} 

\postauthor{ 
	\vspace{0pt} 
	{\par\HorRule} 
	\vspace{-10pt} 
}

\newcommand{\newabstract}[1]{
    {\section*{Abstract}
    \bfseries #1}
  }

\usepackage[authoryear]{natbib}
\title{The Origin of Magnetism in White Dwarfs} 

\shorttitle{Magnetism in White Dwarfs} 

\shortauthors{Ferrario} 

\author{
        \authorstyle{Lilia Ferrario$^1$}
	\newline\newline 
	$^1$\institution{Mathematical Sciences Institute, The
          Australian National University, ACT 2601, Australia}\\ 
      }

\begin{document}

\maketitle 

\thispagestyle{firstpage} 


\newabstract{The absence of magnetic white dwarfs with a
  non-degenerate low-mass stellar companion in a wide binary is still
  very intriguing and at odds with the hypothesis that magnetic white
  dwarfs are the progenies of the magnetically peculiar Ap/Bp
  stars. On the other hand, we cannot resort to a process that
  impedes the generation of a strong magnetic field in the main or
  pre-main sequence progenitors of white dwarfs if they are in a
  multiple stellar system, because such a process would also prevent
  the formation of magnetic cataclysmic variables consisting of a
  magnetic white dwarf accreting mass from a low-mass companion. This
  is the reason why it has been proposed that fields in white dwarfs
  may be linked to their binarity and are generated through a dynamo
  mechanism during common envelope evolution.}


\section{Introduction}

Following the discovery of the first magnetic main sequence star,
78\,Vir \citep{Babcock1947}, highly magnetic compact stars were
predicted to exist under the assumption of magnetic flux conservation
during stellar evolution
\citep[e.g.,][]{Woltjer1964,Angel1981,Tout2004}. The search by
\citet{Preston1970} for Zeeman split lines in the spectra of white
dwarfs (WDs) yielded zero results. Subsequent searches for continuum
polarisation in WDs exhibiting unusual or continuous spectra led to
the discovery of the first MWD, Grw+70$^\circ$\,824
\citep{Kemp1970}. Over the past 50 years hundreds of isolated and
binary MWDs have been discovered \citep[e.g. see][]{Ferrario2015MWD}
and for decades it was assumed that the MWDs were the
descendents of the magnetic main sequence Ap/Bp stars. However, some
recent results on the pairing properties of MWDs have thrown some
doubts on this hypothesis \citep{Tout2008}.

\citet{Kleinman2013} compiled a comprehensive catalogue of nearly
20,000 WDs with SDSS spectra and also identified about
800 magnetic objects while \citet{Kepler2013} provided fields of more
than 500 MWDs.  \citet{Liebert2015} visually searched the 1,735 WD+dM
pairs (nearly 10\% of the WD sample) for detached MWDs+dM, but the
only seemingly detached system they found in the \citet{Kleinman2013}
sample turned out to be the well known polar ST\,LMi
\citep{Ferrario1993STLMi} in a very low state of accretion. Thus, their
search yielded null results, similarly to the study previously conducted by
\citet{Liebert2005} on a much smaller sample of objects. This finding
showed that the hypothesis that magnetic fields in WDs and pairing
with a detached, non-degenerate, low-mass red star are independent is
at the 9$\sigma$ level. This discovery strengthens the hypothesis of
\citet{Regos1995} and \citet{Tout2008} that high field MWDs
($10^6 < B/{\rm G} < 10^ 9$) are generated by a dynamo mechanism
during common envelope evolution that leads to a merging event
\citep[see also][]{Nordhausetal2011,Garciaberro2012}. This hypothesis for
the generation of fields in WDs was investigated further by
\citet{WTF2014}. Population synthesis calculations to explore its
viability were carried out by \citet{Briggs2015} and \citet{Briggs2018MWD}.
\begin{figure}
\centering
\includegraphics[width=0.95\columnwidth]{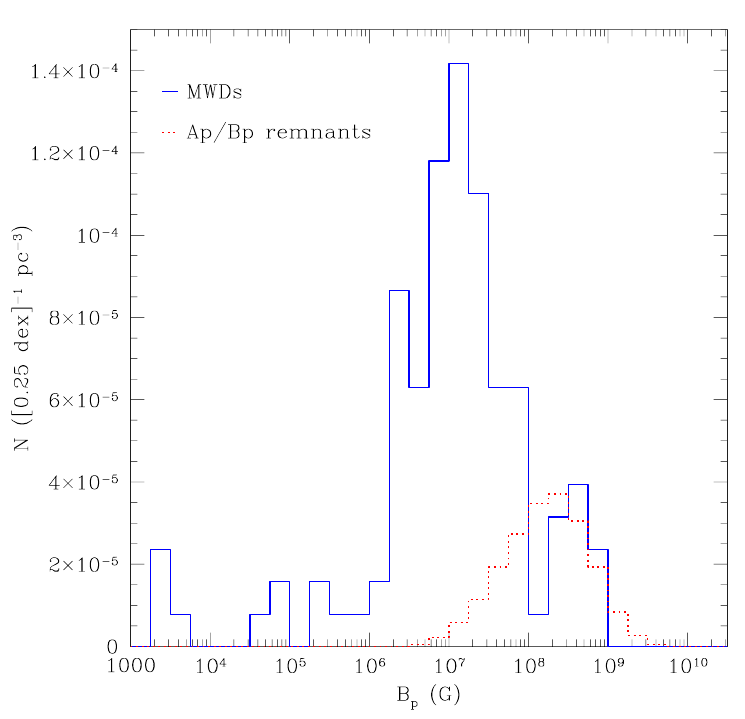}
\caption{Full line: measured mwd incidence. dotted line: predicted
magnetic incidence of Ap/Bp remnants \citep[from][]{Kawka2004}.}
\label{fig:ApBp_fossil}
\end{figure}
In this paper, I will present new results on MWDs and on the origin of
their fields. Comprehensive reviews on magnetic field
generation in stars, from pre-main sequence to the compact star
phase, can be found in \citet{Ferrario2015Origin,Ferrario2018}.

\section{The Ap and Bp stars as the progenitors of the MWDs}

The magnetic main sequence Ap/Bp stars have fields between a few
hundreds G to tens of kG. Originally the estimated birth rate of MWDs
seemed to be compatible with the fossil field hypothesis
\citep{Angel1981} that predicts Ap/Bp stars to evolve into MWDs under
magnetic flux conservation. However, more recent studies by
\citet{Kawka2004} showed that the birth rates and fields of Ap/Bp
stars are not consistent with those of MWDs under flux
conservation. Fig.\,\ref{fig:ApBp_fossil} shows the observed field
distribution normalised to the observed space density of MWDs compared
to the predicted field distribution of magnetic Ap/Bp remnants
normalised to the predicted space density
\citep[from][]{Kawka2004}. Thus, the fields of Ap/Bp stars (i) only
map onto the highest field MWDs and (ii) the birthrate of Ap/Bp is not
sufficiently high to explain the incidence of magnetism among
WDs. This inconsistency was also highlighted by the work of
\citet{WF2005}.  In order to make the fossil field hypothesis remain a
viable option to explain the origin of magnetic fields in WDs,
\citet{WF2005} proposed a model where 40\% of main sequence stars with
$M >4.5$\,M$_\odot$ are magnetic with dipolar fields of 10-100\,G, a
field regime that had not yet been investigated. However, their
assumption was not corroborated by observations performed in following
years with new state-of-the-art spectropolarimeters
\citep{Auriere2007}. These surveys not only ruled out the existence of
a population of weakly magnetic Ap/Bp stars but also revealed the
presence of a ``magnetic desert'' between 300\,G (which turned out to
be the lower bound of magnetism in Ap/Bp stars) and the detection
limit of the surveys. Thus, this discovery revealed that magnetic
Ap/Bp stars are either magnetic at the 300\,G level or are not
magnetic at all \citep{Auriere2007}.

These findings leave a sizeable fraction of MWDs at the low-field end
of the distribution ($\sim10^6$ to a few $10^7$\,G) without obvious
progenitors of the magnetic Ap/Bp class. Furthermore, the studies of
\citet{WF2005} also disclosed that the higher than average
mass of MWDs \citep{Liebert1988,Ferrario2015MWD} cannot be explained via
the Ap/Bp progenitor scenario unless either the initial to final mass
function of WDs is significantly altered by the presence of a magnetic
field or an hitherto undetected population of massive and weakly
magnetic main sequence stars can evolve into MWD. The latter
hypothesis, however, seems to be unlikely given the observational
evidence discussed above.

\section{The common envelope hypothesis for the origin of MWDs}

\citet{Briggs2015} and \citet{Briggs2018MWD} conducted population
synthesis calculations to verify the validity of the common envelope
merging scenario hypothesis of \citet{Tout2008}. In this scenario the
magnetic field is generated from the differential rotation of cores
that spiral in toward each other during common envelope evolution.
The closer the cores get before the envelope is ejected the stronger
the field that is produced. Thus, the isolated MWDs arise when stars
merge while the next strongest fields are generated in binaries that
come out of common envelope on close orbits and will evolve into
magnetic cataclysmic variables (MCVs) which are composed of a MWD
accreting matter from a low-mass companion. The fields in MCVs are
measured via cyclotron and Zeeman spectroscopy and are in the range
$\sim 10^7-10^8$\,G \citep[see, e.g.,][]{Ferrario1992,Ferrario1996}
for the more strongly magnetic ``polars'' and $\sim 10^6-10^7$\,G in
the more weakly magnetic ``intermediate polars''
\citep[e.g. see][]{FWK1993,FW1993}. The existence of MCVs highlights
the fact that the pairing of MWDs with low-mass red dwarf stars is
indeed possible, what is puzzling is the absence of wide,
non-interacting binaries of this kind even if such a pairing is very
common among non-magnetic WDs.

\citet{Tout2008} proposed that the low-accretion rate polars (LARPS),
whose stellar components are close enough to allow the WD to capture a
weak stellar wind from its low-mass companion, could be pre-polars
waiting for gravitational radiation to bring the stars sufficiently
close to allow Roche lobe overflow \citep[see also][who renamed these
systems pre-polars or ``PREPs'' to differentiate them from polars in a
low-state of accretion]{Schwope2009}. Although this scenario for the
origin of fields in WDs is very attractive, \citet{PotterTout2010}
found that the time-scale for the diffusion of the field into the WD
is much longer than the presumed lifetime of the common envelope.
Therefore \citet{WTF2014} proposed that during common envelope
evolution a weak seed poloidal field deeply anchored in the WD gets
wound up by differential rotation in the envelope of the merged object
that will evolve into a WD. In this model, the dynamo action would
amplify the poloidal field until a stable poloidal/toroidal structure
is achieved.

\begin{figure}
\centering
\includegraphics[width=1.0\columnwidth]{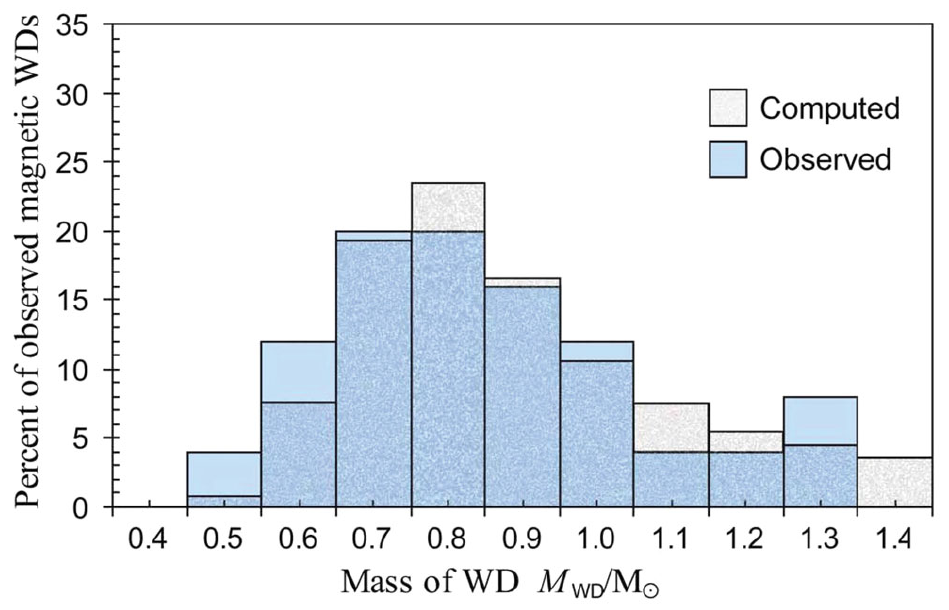}
\caption{Observed masses of high field MWDs compared with the
  theoretical sample \citep{Briggs2015}.}
\label{fig:MassMWDs}
\end{figure}

The population synthesis calculations of \citet{Briggs2015} have shown
that the common envelope hypothesis successfully explains the mass distribution
of MWDs (see Fig.\,\ref{fig:MassMWDs}). They found that the major
contribution to the observed population of MWDs comes from the
degenerate core of AGB stars merging with low-mass main-sequence
stars.  Merging events of a WD with another WD also occur but at a much lower
rate and the resulting objects occupy the high-mass end of the MWD
distribution.
\citet{Briggs2018MWD} extended their studies to the field
distribution of MWDs and assigned a field strength to each member of
their synthetic sample on the basis of the dynamo model results of
\citet{WTF2014}. According to this model the field strength achieved
by the WD during common envelope evolution is proportional to the
orbital angular velocity
$\Omega=\displaystyle{\frac{2\pi}{P_{\rm orb}}}$ of the system when
the envelope is ejected. In this simple picture the object emerging
from common envelope has a field
\begin{equation}\label{Field}
B = B_0\left(\frac{\Omega}{\Omega_{\rm crit}}\right)\, \mbox{G}.
\end{equation}
where $\Omega_{\rm crit}$ is the break-up angular velocity of the WD
and the parameter $B_0$ is determined empirically to best map the
theoretical field range to the observed one. This means that different a
$B_0$ shifts the field distribution to lower or higher fields. The
actual shape of the field distribution is given by the common envelope
efficiency parameter $\alpha$ which they found to be less than about
0.3 \citep[see][for further details on the modelling]{Briggs2015,
  Briggs2018MWD}.

Because the field $B$ in equation (\ref{Field}) is inversely
proportional to the orbital period, $P_{\rm orb}$,
\citet{Briggs2018MWD} find that low-mass MWDs arising from the merging
of the degenerate core of an RGB star with a very-low mass convective
star (CS) gives rise to fields generally stronger than those predicted
for more massive MWDs created by the degenerate core of an AGB star
merging with a main sequence star. The merging of two WDs are not part
of this prediction but because $\Omega_{\rm crit}$ can only be reached
during the merging of a very compact double WD system (double
degenerate ``DD'' channel), such mergers are envisaged to produce MWDs
that are strongly magnetic, massive, and rapidly spinning. The MWD
RE\,J0317-853 \citep{Barstow1995,Ferrario1997,Vennes2003} is certainly
a good DD merger candidate. So far, there are not many MWDs for which
mass and field are both known, so it is not possible to verify this
theoretical prediction. Nonetheless, observations seem to indicate
that the currently known ultra-massive MWDs do have fields at the low
end of the distribution \citep[e.g., 1RXS\,J0823.6-2525,
PG\,1658+441,][]{Ferrario1998,Schmidt2001}.

\citet{Ferrario2012} shows that the combination of stellar components
that can best fit all current observational constraints of wide binary
systems yields 18\% of WDs paired with an M dwarf, 47\% of Sirius-type
systems and 35\% of non-interacting WD-WD systems \citep[that is,
objects such as PG\,1346+082,][are excluded]{Provencal1997}. We
already know that there are no MWDs paired with M dwarfs in a wide
binary. The studies of \citet{Rolland2015} have revealed that more
than 60\% of objects in their sample must be in wide binaries composed
of either a MWD and a featureless (DC) WD or a MWD and a hydrogen (DA)
WD.  \citet{Kawka2017} list all non-interacting double WD systems in
which one of the two components is magnetic. Some of these are very
wide common proper motion systems so that the two stars could not have
interacted during their evolution. Interestingly, some of these
binaries show that the ages of the two components are inconsistent if
single star evolution is assumed. Thus \citet{Kawka2017} suggest that
they might have been triple systems composed of two stars that merged
during common envelope evolution and a third star that never
interacted with the other two.  They also show that the magnetic field
of one of the two components of the DD system NLTT\,12758 may have
arisen during CE evolution, in a manner similar to that proposed by
\citet{Briggs2018MCV} for MCVs. In summary, it would be interesting to
further investigate the characteristics of magnetic DD systems to establish
whether the shorter period systems may all have formed via binary interaction
while the longer period (and thus wider) systems may have initially been
triple systems. If neither of these channels can be invoked to explain
the existence of at least some of these magnetic DD systems, the possibility of
these arising from single star evolution (the Ap/Bp channel) remains open.

The detection and thus study of a hitherto unknown population of MWDs hidden 
in the glare of brighter companions (spectral type K or earlier) may
be of a much more challenging nature.

 \begin{figure}
\centering
\includegraphics[width=1.0\columnwidth]{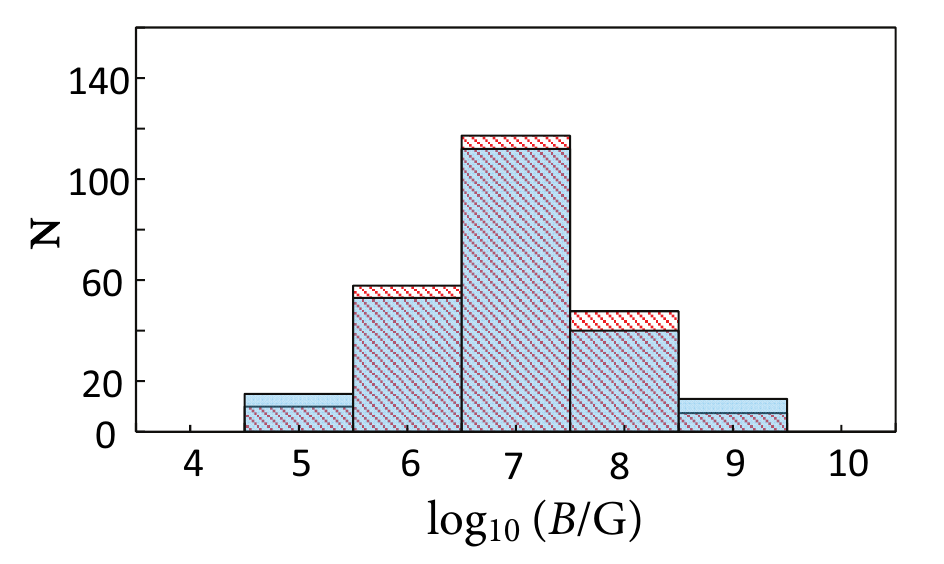}
\caption{Observed high field MWDs  (blue histogram) compared with the theoretical sample (pink histogram)
  \citep{Briggs2018MWD}.}
\label{fig:Bmag}
\end{figure}
\subsection{Cool magnetic white dwarfs}

About 30\,per cent of WD spectra show evidence for Ca, Si, Mg, Fe, Na
and other metals \citep{Zuckerman2003} and about 13 to 23\% of cool
($T_{\rm eff}<8\,000$\,K) metal polluted WDs are magnetic
\citep{Kawka2014, Kawka2018, Hollands2015,Hollands2017}. Thus, the
incidence of magnetism among cool polluted WDs is much higher than
among ordinary WDs \citep[about 3\, percent][]{Ferrario2015MWD}. It is
tempting to hypothesise that the generally weak magnetic fields
observed in these cool and polluted WDs
\citep[$0.1\le B/10^7{\rm G}\le 1.1$,][]{Hollands2017, Kawka2018} may
be caused by a WD accreting a gaseous giant planet (that would
generate their weak fields) and other rocky debris that would pollute
their atmospheres.  The engulfment of giant planets and rocky debris
could occur during the AGB evolution when planets and other minor
bodies drift toward the degenerate stellar core owing to frictional
forces as they moves through the envelope of the star \citep[e.g.][]{
  Li1998}. However, if magnetism (and possibly pollution) arise early
in the life of a WD, then one would expect that hot (and possibly
polluted) WDs would exhibit the same incidence of magnetism as cool
and polluted WDs, but this does not seem to be the
case. \citet{Farihi2011} propose a different mechanism involving close
stellar encounters that would perturb the orbits of outer bodies such
as large gaseous planets and asteroid belts and cause their inward
migration and subsequent accretion by the WD. Over a cooling age of
$2-9$\,billion years, this kind of stellar encounters is not unlikely
to occur and could account for the existence of the cool and polluted
MWD G77--50 \citep{Farihi2011}. This mechanism may also explain the
high incidence of magnetism among cool white dwarfs
\citep{Liebert1979, Fabrika1999, Kawka2007}. Although this result has
not been fully corroborated by observations yet, future observations
of a larger sample of cool MWDs may shed some light on this issue and
will allow us to accept or reject the possibility that weak fields
could be generated via the accretion of giant gaseous planets
\citep{Kawka2018NLTT754}.
  
\section*{Acknowledgements}

I wish to thank St\'ephane Vennes, Adela Kawka and John Landstreet for
stimulating discussions.


\bibliography{Ferrario}

\end{document}